\documentstyle[12pt]{article}

\title{
{\vspace{-3cm} \normalsize
\hfill \parbox{27mm}{DESY 94-006 }\\
\hfill \parbox{54mm}{ITP-Budapest Rep. No. 505  }\\
\hfill \parbox{35mm}{hep-ph/9401290}}\\[25mm]
Grand unification for mirror fermions  }

\author{
 F. Csikor \\[4mm]
Deutsches Elektronen-Synchrotron DESY, \\
Notkestr.\,85, D-22603 Hamburg, FRG \\
and \\
 Institute for Theoretical Physics, E\"otv\"os University,
 Budapest \thanks{Permanent address}      \\[6mm]
I. Montvay  \\[4mm]
Deutsches Elektronen-Synchrotron DESY, \\
Notkestr.\,85, D-22603 Hamburg, FRG }

\date{January, 1994}

\newcommand{\be}{\begin{equation}}
\newcommand{\ee}{\end{equation}}
\newcommand{\bd}{\begin{displaymath}}
\newcommand{\ed}{\end{displaymath}}
\newcommand{\bea}{\begin{eqnarray}}
\newcommand{\eea}{\end{eqnarray}}
\newcommand{\GeV}{\mbox{\rm \, GeV}}

\begin{document}
\maketitle

\begin{abstract} \normalsize
 The possibility of grand unification of the standard model (SM) with
 fermion spectrum extended to include mirror fermions is examined.
 SM gauge couplings do not automatically unify.
 SO(10) grand unification is studied with one intermediate scale.
 Renormalization group equations (RGE) for fermion Yukawa couplings and
 the scalar self-coupling are studied numerically at one and two loop
 level.
 Strong restrictions for mirror fermion masses are obtained assuming
 perturbative unification.
 Mirror masses much smaller than the tree unitarity bounds are
 required.
 In particular mirror leptons have to be around 50 GeV.
 Consistency of the mirror fermion model with LEP precision data is
 established.
 A direct search for single production of mirror neutrinos at LEP could
 exclude or confirm the GUT version of the mirror fermion model.
\end{abstract}

\newpage   \vspace*{1cm}
\section { Introduction }
 In this note we study the possibilities of perturbative grand
 unification for the mirror fermion extension of the SM.
 This very simple extension of the SM \cite{l1} is to enlarge only
 the fermion content by introducing mirror (i.e. opposite chirality
 property) fermions to each fermion of the SM (i.e. to each ordinary
 fermion), preserving the $\rm SU(3) \otimes SU(2) \otimes U(1)$
 group structure.
 Ordinary and mirror fermions are allowed to mix.
 In fact mixing is necessary in order to avoid stable mirror fermions.
 Present experiments directly exclude mirror fermions with masses
 below roughly half of the $Z^0$ mass.
 Heavier mirror fermions are still allowed.
 Many of the  phenomenological consequences of such a model have been
 worked out in \cite{l2}, \cite{l3}, and motivations were summarized
 in \cite{l3a}.
 (For a review of earlier work on models with mirror fermions see
 \cite{l6}.)

 The mixing angles of ordinary and mirror fermions are small.
 Constraints from experimental data have been worked out in \cite{l7}.
 Mixing angles are typically bounded by 0.1 - 0.2.
 Even more resrictive upper bounds (0.02)
 for the leptonic mixing angles were obtained in \cite{l4}.
 Also a  recent fit to LEP precision data shows that the model
 is  consistent with experiment for small (but non-zero)
 mixing angles \cite{l5}.
 Since ordinary (left or right) and mirror fermions (right or left,
 respectively) transform identically under the
 $\rm SU(3) \otimes SU(2) \otimes U(1)$ group it is possible to write
 down invariant mixing mass terms.
 Such mass terms of the order of the symmetry breaking scale would
 imply large mixings, therefore small mixing angles are imposed as
 an experimental constraint.
 In GUT it is easy to forbid the mixing mass terms
 invoking discrete symmetries, therefore in the following we
 neglect mixing effects.

 We emphasize that we are concerned with the unification of
 the above described simple mirror fermion model assuming minimal
 extension of the standard model.
 Thus in our case mirror fermions get a mass at the scale of
 electroweak symmetry breaking.

\section{ Gauge coupling unification }
 It is well known that starting with the measured values of the
 gauge couplings, the running couplings do not meet at a single scale
 \cite{l8} in the SM.
 So it is interesting to examine whether mirror fermions change the
 situation.
 Since (by assumption) each ordinary fermion has a
 mirror partner of similar quantum numbers, it is easy to see
 that at one loop the slope of the running couplings gets the
 same contribution for all the three gauge couplings.
 Thus only the actual values of the couplings are modified,
 but they do not meet, similarly to the SM case.
 We have checked that neither threshold
 effects nor two loop RG effects change this conclusion.

 Clearly, we may have GUT only with symmetry breaking in at
 least two steps.
 We have choosen SO(10) as the grand unifying group.
 Among other attractive features \cite{l9} we want to keep
 the number of new fermions small.
 Assuming two step symmetry breaking
 the intermediate scale as well as the GUT scale is determined by
 the too loop RG equations and low energy experimental input
 couplings \cite{l10}.
 Using the notation of \cite{l10} we have
\be
\frac{d\omega_i (\mu )}{d\ln \mu }=-\frac{a_i }{2\pi }-\sum_j
\frac{b_{ij}}{8\pi^2 \omega _j},
\ee
 where
\be
\omega_i =\alpha_i ^{-1} =4\pi /g_i ^2 \, .
\ee

 The initial values are:
 $\alpha_1 (M_Z) = 0.016887 \pm 0.000040$,
 $\alpha_2 (M_Z) = 0.03322 \pm 0.00025$,
 $\alpha_3 (M_Z) = 0.120 \pm 0.008$.
 Between $M_Z$ and $M_I$ (the intermediate scale) the constants
 are given by
\be
a=\left( \begin{array}{c}\frac{81}{10}\\[1em]
                        -\frac{5}{6}\\[1em] -3
\end{array} \right) \: ,
\hspace{2em}
b=\left( \begin{array}{ccc}
\frac{1149}{50}&\frac{9}{2}&\frac{264}{15}\\[1em]
\frac{3}{2}&\frac{329}{6}&24\\[1em]
\frac{11}{5}&9&-20
\end{array} \right) \: .
\ee
 Between $M_I$ and $M_U$ (unification scale) the constants
 $a_i$ and $b_{ij}$ depend on the intermediate range unbroken gauge
 group $G_I$ as well as the Higgs multiplets remaining massless
 at the scale $M_I$.
 We consider all the possibilities listed in \cite{l10}
 i.e. $G_I$=\{$2_L 2_R 4_C $\},
    \{$2_L 2_R 4_C \otimes P$\},
    \{$2_L 2_R 1_X 3_c $\},
    \{$2_L 2_R 1_X 3_c \otimes P$\},
    \{$2_L 1_R 4_C $\},
    \{$2_L 1_R 1_X 3_c $\}.
 \{$2_L 2_R 4_C $\} e.g.\ stands here for the group
 $\rm SU(2)_L \otimes  SU(2)_R \otimes SU(4)_C $.
 $X=(B-L)/2$ and the factor $P$ is an unbroken parity symmetry.
 The Higgs content does not change by considering the mirror fermion
 model so it is the same as given in Table I of  \cite{l10}.
 The $a_i$'s all increase by 4 as compared to
 \cite{l10}, the change of the $b_{ij}$'s is more complicated.
 To save space we do not reproduce the actual values here.

 Solving the RG equations and applying the appropriate (2-loop)
 matching conditions we arrive at the scales and couplings at $M_U$
 as listed in Table I.
 There is no solution for $G_I = \{2_L 1_R 1_X 3_c \}$.
 Not all these solutions are acceptable, because of the constraint
 coming from the experimental lower limit on the proton lifetime.
 Chains 1a, 2a survive, 1b and 2b are marginal, while
 chain 3 is definitely ruled out.

\section{ Scalar couplings }
\subsection{ RG equations for the Yukawa couplings }
 As well known the running of Yukawa couplings suffers from
 Landau poles in the one loop approximation.
 The presence of these singularities at some scale $\Lambda_{Yuk.}$
 highlights the breakdown of perturbation theory and the probable
 triviality of the continuum limit.
 It is reasonable to accept as perturbative regions those
 scales, where the squared Yukawa couplings are less than $4\pi$,
 (i.e.\ $\alpha _{Yuk.}  \leq 1$.)
 The RG equations for the Yukawa couplings are given
 in a general gauge theory in \cite{l11} for both the one and two
 loop case.
 The initial values are given by relating fermion masses to the
 Yukawa couplings at threshold.
 Even though the fermion mass spectrum is not known
 (top and mirror fermion masses are unknown), we start the Yukawa
 coupling evolutions from thresholds assuming 'reasonable' masses.
 Useful guides on the mirror fermion masses are the experimental lower
 bounds and also the tree unitarity upper bounds derived in \cite{l3}.
 Mirror doublets are always assumed to be degenerate as required
 to reproduce the precision LEP data (\cite{l5}).
 Moreover it is natural to assume that mirror fermions are always
 heavier than the corresponding ordinary fermions (this is non trivial
 for the mirror top only.)
 For the top quark mass we assume values consistent with the fit of
 precision LEP data.
 The other condition on the running Yukawa couplings is that
 $\alpha_{Yuk.} $ should remain less than one below $M_I$,
 (i.e. $\Lambda_{Yuk.} \geq M_I$).
 The latter is determined from the gauge coupling RG equations
 assuming one of the possible two step symmetry breakings of SO(10) as
 explained above.

 We have solved the one loop RG equations numerically for many
 representative choices of masses.
 We found that mirror leptons should be light, much lighter than
 mirror quarks in order to get 'reasonably' high masses at all.
 This conclusion is in accord with \cite{l12}, where Yukawa coupling
 evolution is studied starting from a high scale.
 Namely, the (mathematical) infrared fixed point is reached at
 vanishing lepton masses.
 Also the mirror quark masses should be relatively small.
 This is again consistent with \cite{l12}, where an upper bound on
 the sum of quark mass squares is derived.
 Some numerical examples are:
 $M_{mirror} = 92 \GeV$, $M_{top} = 150 \GeV$
 yields $\Lambda_{Yuk.} = 10^{10.26} \GeV$; $M_{m.lepton} = 50 \GeV$,
 $M_{m.quark} = 92 \GeV$, $M_{top} = 150 \GeV$ yields
 $\Lambda_{Yuk.} = 10^{22.03} \GeV$; $M_{m.lepton} = 50 \GeV$,
 $M_{m.quark} = 92 \GeV$,
 $M_{m.top} = M_{m.bottom} = M_{top} = 145 \GeV$ yields
 $\Lambda_{Yuk.} = 10^{11.1} \GeV$.
 Compared to the tree unitarity upper bounds of \cite{l3} these
 values of mirror fermion masses are rather small.
 In particular in \cite{l5} also higher masses in the range
 (100-300) GeV have been assumed.

\subsection{ RG equation for the scalar quartic self coupling}
 Already at one loop we have a coupled system of differential
 equations for the gauge, Yukawa and quartic couplings.
 The RG equations are given in the general case for one and two
 loop level in \cite{l11} and \cite{l13}.
 Though the Higgs mass is unknown, we continue our practice to assume
 a 'reasonable' value to provide an initial condition to the RG
 equation.
 Following \cite{l14} the RG evolution of the scalar quartic coupling
 may be used to establish upper and lower bounds on the Higgs mass.
 The lower bound arises from requiring a positive quartic coupling
 ($\lambda $).
 The upper bound arises from the Landau pole of $\lambda $
 (triviality bound).
 The corresponding scales are:
 $\Lambda_{inst.} $ and $\Lambda_ \lambda $.
 In general
 $\Lambda_{inst.} \leq \Lambda_ \lambda \leq \Lambda_{Yuk.} $
 for given fermion masses.
 Combining these with the information obtained from gauge coupling
 evolution for $M_I$, we have the condition
 $M_I \leq \Lambda_{inst.} $,
 i.e.\ only sufficiently high $\Lambda_{inst.} $ is acceptable.

 Solving the RG equations numerically we find that depending on
 the choice of fermion masses the allowed range of the
 Higgs mass is rather restricted or even empty.
 Nevertheless for reasonably light mirror leptons and mirror quarks
 the perturbative region may extend to $M_X$.
 Some numerical examples are: for
 $M_{mirror} = 92 \GeV$, $M_{top} = 150 \GeV$,
 ($\Lambda_{Yuk.} = 10^{10.26} \GeV$),
 $M_{Higgs} \in (222.6 \GeV,222.9 \GeV)$  is  acceptable
 for chain 1a and $M_{Higgs} \in (221.9 \GeV,225.5 \GeV)$ is acceptable
 for chain 2a.
 For $M_{m.lepton} = 50 \GeV$, $M_{m.quark} = 92 \GeV$,
 $M_{top} = 150 \GeV$,
 ($\Lambda_{Yuk.} = 10^{22.03} \GeV$)
 $M_{Higgs} \in (193 \GeV,227 \GeV)$ is acceptable
 for chain 1a and $M_{Higgs} \in (190 \GeV,238 \GeV)$ for chain 2a.
 For $M_{m.lepton} = 50 \GeV$, $M_{m.quark} = 92 \GeV$,
 $M_{m.top} = M_{m.bottom} = M_{top} = 145 \GeV$
 ($\Lambda_{Yuk.} = 10^{11.1} \GeV$)
 $M_{Higgs} \in (240.63 \GeV,242.5 \GeV)$ is acceptable
 for chain 1a and $M_{Higgs} \in (239.8 \GeV,246 \GeV)$ for chain 2a.

\subsection{ Two loop RG effects for scalar couplings }
 Using  the complete two loop RG equations, besides the infrared
 fixed point, the possibility of a second (ultraviolet) fixed point
 arises.
 Instead of trying to solve the nonlinear equations determining the
 second fixed point, we have solved the RG equations numerically and
 observed the peculiar scale dependence of the couplings.
 Namely, below the second fixed point the scalar couplings are almost
 constant (this corresponds to the infrared fixed point)
 and after a short transition period the couplings are again
 constant at different values.
 The transition scale is near the Landau pole of the one loop RG
 equations.
 The other possibility is that new ultraviolet fixed points do
 not arise, so the Landau pole does not disappear.
 An important question to answer is, whether or not the
 fixed point behaviour belongs to the perturbative range or not, i.e.\
 do couplings remain sufficiently small (so that e.g.\ the
 'fine structure constants' associated to the different
 couplings are all less than unity.)
 We find that this does not happen.
 The perturbative regime does not appreciably change when the
 one loop RG equations are replaced by the two loop ones.
 It follows that the above one loop results on the allowed range
 of mirror fermion and Higgs masses will not be changed by the more
 sophisticated two loop treatment, provided that perturbative
 unification is  assumed.
 An example of a simplified model is shown in figs.\ 1,2.
 We have kept only the top and mirror quarks, the Higgs  and
 the QCD coupling.
 The masses are
 $M_{top}= 150 \GeV$, $M_{m.quark} = 92 \GeV$ and
 $M_{Higgs} = 300 \GeV$.

\section{ Constraints implied by LEP data }
\subsection{ Precision data }
 At low energies the really crucial test of any theory
 beyond the SM is whether it survives a comparison with LEP
 precision data.
 For the mirror fermion model such a comparison has been performed
 in \cite{l5}.
 Since that paper assumed somewhat higher mirror masses than
 allowed by perturbative unification the analyses has to
 be repeated with lower mirror masses.
 The fitted parameters are quark mixing angles ($\alpha_q$)
 assumed to be equal for u, c quarks (those of d and s quarks are
 not free parameters) and the top quark mixing angle ($\alpha_{top}$)
 (the bottom quark mixing angle is not a free paramater).
 We have used preliminary 1992 data as given in \cite{l15}.
 In \cite{l5} it was found that zero mixing angles are
 already excluded, but for suitable mixing angles very good fits
 to LEP and low energy neutrino data are obtained.
 This qualitative statement remains unchanged for the lower
 mirror masses as well.
 An example of the equal $\chi ^2 $ curves is given in fig.\ 3 for the
 following input parameters:
 $M_{m.lepton} = 50 \GeV$, $M_{m.quark} = 92 \GeV$,
 $M_{top} = M_{m.top} = M_{m.bottom} = 150 \GeV$,
 $M_{Higgs} = 250 \GeV$, $\alpha_s = 0.12$ and the right leptonic
 mixing angles are zero, the left leptonic mixing angles
 are equal to 0.092.
 Good fits for lower values of the left leptonic mixing angles
 are possible at the expense of increasing the top mass.

\subsection{ Direct search for single production}
 The pair production of mirror fermions is excluded by experimental
 data.
 However, single production through the mixing vertex coupling
 ordinary and mirror fermions to the weak vector bosons is also
 possible.
 (The same vertex is responsible for the decay, which goes into an
 ordinary fermion and a possibly virtual vector boson.
 Decay to ordinary lepton and a photon occurs only in second
 order and is very small.)
 The cross-section depends on the mixing angles (it is given
 e.g.\ in \cite{l20}).
 Therefore experiment will give combined upper limits on mirror
 fermion masses and mixings.
 To our knowledge such a search has not been performed so far.
 The L3 search for singly produced excited neutrinos decaying to
 $eW$ \cite{l21} can be used to derive a rough upper bound on
 the mixing angles as a function of mirror neutrino mass.
 (The estimate is very rough, since the angular distribution
 of mirror neutrino and excited neutrino is very different.)
 For $M_{m.lepton} = 50 \GeV$, zero right leptonic mixing angles
 and equal left leptonic mixing angles, we got an upper bound
 of 0.054 for the latter.
 Thus, a systematic search of single production of mirror
 neutrinos combined with other LEP precision data may easily
 exclude the low masses required by the GUT scenario.

\section{ Conclusion }
 Assuming perturbative grand unification for the mirror fermion
 extension of the standard model, we find that intermediate
 scale symmetry breaking is necessary.
 Assuming two step symmetry breaking the intermediate scale can be
 determined from gauge coupling evolution.
 Mirror fermion masses are severely restricted by the requirement
 that Yukawa couplings should remain small during the evolution
 below the intermediate scale.
 Quite restrictive information on the Higgs mass is obtained
 assuming positive and small quartic coupling during evolution.
 Mirror lepton masses turn out to be small (near half of the $Z$ mass),
 mirror quark masses should be also much smaller
 than allowed by the tree unitarity bounds.
 For suitable mixing angles LEP precision data can be very well
 fitted with the mirror masses allowed by the above considerations.
 A combined study of LEP precision data with a direct search for single
 mirror neutrino production could easily exclude the low mirror
 lepton masses required by perturbative grand unification.

\vspace{1cm}
\large\bf Acknowledgement \normalsize\rm
\vspace{1em}

 One of us (F.Cs.) thanks the hospitality of the DESY Theory group
 during the preparation of the paper and acknowledges partial
 support from Hung. Sci. Grant under Contract No. OTKA-I/3-2190.

%%%%%%%%%%%%%%%%%%%%%%%%%%%%%%%%%%%%%%%%%%%%%%%%%%%%%%%%%%%%%%%%%%%%

%%%%%%%%%%%%%%%%%%%%%%%%%%%%%%%%%%%%%%%%%%%%%%%%%%%%%%%%%%%%%%%%%%
\newpage
\vspace{2cm}
\begin{center}     \Large\bf Table caption   \rm\normalsize
\end{center}
\vspace{1em}

\bf Table \, 1.   \hspace{5pt} \rm
 Intermediate scale ($M_I$) and unification scale ($M_U$)
 obtained by solving the renormalization group equations
 for different intermediate symmetry groups ($G_I$).

\vspace{3cm}

\begin{center}     \Large\bf Table 1    \rm\normalsize
\end{center}

\begin{center}
\begin{tabular}{|c |c|c|c|c|} \hline

Chain & $G_I$ & \multicolumn{2}{c|}{$\log _{10} (M/1\, \GeV )$} &
 $\omega _U $ \\
\cline{3-4}
      &       & $M_I$ & $M_U$                      &     \\
\hline
1a & $2_L 2_R 4_C $& 10.11 & 16.32 & 23.58 \\
1b & $2_L 2_R 4_C \otimes P$& 13.70 & 14.85 & 21.01 \\
2a & $2_L 2_R 1_X 3_c $ & 9.35 & 16.37 & 23.67 \\
2b & $2_L 2_R 1_X 3_c \otimes P$ & 10.66 & 15.34 & 22.94 \\
3 & $2_L 1_R 4_C $ & 11.30 & 14.40 & 24.87 \\
\hline
\end{tabular}
\end{center}

%%%%%%%%%%%%%%%%%%%%%%%%%%%%%%%%%%%%%%%%%%%%%%%%%%%%%%%%%%%%%%%%%%

\newpage
\vspace{2cm}
\begin{center}  \Large\bf Figure captions \normalsize\rm
\end{center}
\vspace{1em}

\bf Fig.\,1. \hspace{5pt} \rm
 The running couplings as a function of $t = \log _{10} (\Lambda )$
 at one loop order in  the simplified model containing
 only the top quark and mirror quarks, Higgs  and the QCD coupling.
 The input masses are
 $M_{top} = 150 \GeV$, $M_{m.quark} = 92 \GeV$
 and $M_{Higgs} = 300 \GeV$.
\vspace{20pt}

\bf Fig.\,2. \hspace{5pt} \rm
 The same as fig.\ 1 at two loop order.
\vspace{20pt}

\bf Fig.\,3. \hspace{5pt} \rm
 Equal $\chi ^2 $ curves of a fit to LEP and low energy data.
 The area between the indicated curves belongs to
 the lowest $\chi ^2 $.
 The other curves correspond to $\chi ^2 $'s increasing
 by steps of 0.5.

\end{document}